\documentclass[pre,aps,twocolumn,showpacs,showkeys,preprintnumbers,longbibliography,superscriptaddress,amsmath,amssymb,floatfix,nofootinbib,a4paper]{revtex4-2}

\usepackage{graphicx}
\usepackage[table,usenames,svgnames]{xcolor}
\usepackage{multirow}
\usepackage{natbib}
\usepackage{bm}

\newcommand{\br}{{\bf r}}

\newcommand{\bj}{{\bf j}}
\newcommand{\be}{{\bf e}}

\newcommand{\bk}{{\bf k}}

\newcommand{\bu}{{\bf u}}

\definecolor{myred}{rgb}{0.93,0.27,0.27}
\definecolor{myblue}{rgb}{0.4,0.6,1}
\definecolor{mygreen}{rgb}{0.14,0.8,0.14}

\newcommand{\dens}{n}
\newcommand{\dn}{n^\mathrm{(dev)}}
\newcommand{\ns}{\dens_\mathrm{s}}
\newcommand{\nb}{\dens_\mathrm{b}}
\newcommand{\nhvar}[1]{{\hat{\dens}_\mathrm{#1}}}
\newcommand{\nhs}{{\nhvar{s}}}
\newcommand{\nhb}{{\nhvar{b}}}
\newcommand{\Dvar}[1]{{D_\mathrm{#1}}}
\newcommand{\Ds}{{\Dvar{s}}}
\newcommand{\Db}{{\Dvar{b}}}
\newcommand{\Ls}{{L_\mathrm{s}}}
\newcommand{\Lb}{{L_\mathrm{b}}}
\newcommand{\pec}{\mathrm{Pe}}
\newcommand{\Peb}{\pec_\mathrm{b}}
\newcommand{\Pes}{\pec_\mathrm{s}}
\newcommand{\phib}{\phi_\mathrm{b}}
\newcommand{\phis}{\phi_\mathrm{s}}
\newcommand{\rb}{R_\mathrm{b}}
\newcommand{\rs}{R_\mathrm{s}}

\newcommand{\eq}[1]{Eq.~(\ref{#1})}
\newcommand{\eqs}[1]{Eqs.~(\ref{#1})}
\newcommand{\rcite}[1]{Ref.~\cite{#1}}
\newcommand{\rcites}[1]{Refs.~\cite{#1}}
\newcommand{\fref}[1]{Fig.~\ref{#1}}
\newcommand{\sref}[1]{Sec.~\ref{#1}}

\newcommand{\Fref}[1]{Figure~\ref{#1}}

\begin{document}

\title{Hydrodynamic interactions in a binary--mixture
  colloidal monolayer}

\author{M.~Chamorro-Burgos}
\affiliation{F\'\i sica Te\'orica, Universidad de Sevilla, Apdo.~1065, 
  41080 Sevilla, Spain}

\author{Alvaro Dom\'\i nguez}
\email{\texttt{dominguez@us.es}}
\affiliation{F\'\i sica Te\'orica, Universidad de Sevilla, Apdo.~1065, 
  41080 Sevilla, Spain}
\affiliation{Instituto Carlos I de F{\'i}sica Te{\'o}rica y  Computacional, 18071 Granada, Spain}

\date{received 12 January 2026; accepted 7 April 2026; published 4 May 2026}

\begin{abstract}
  A colloidal monolayer embedded in the bulk of a fluid experiences a
  ``compressible'', long-range hydrodynamic interaction which, far
  from boundaries, leads to a breakdown of Fick's law above a well
  defined length scale, showing up as anomalous collective
  diffusion. We here extend the model to study the effect of the
  hydrodynamic interaction on a monolayer formed by two types of
  particles. The most interesting finding is a new regime, in the
  limit of very dissimilar kinds of particles, where the effective
  dynamics of the concentration of ``big'' (slow) particles appears to
  obey Fick's law at large scales, but the corresponding collective
  diffusivity is completely determined, through hydrodynamic coupling,
  by the diffusivity of the ``small'' (fast) particles.
\end{abstract}

\maketitle

\section{Introduction}
\label{sec:intro}

\begin{figure}[t]
  \hfill
  \begin{tabular}[c]{c}
    side view
    \\
    \includegraphics[width=.25\textwidth]{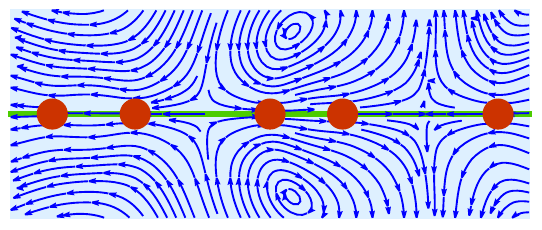}
  \end{tabular} \hfill
  \begin{tabular}[c]{c}
    top view
    \\
    \includegraphics[width=.15\textwidth]{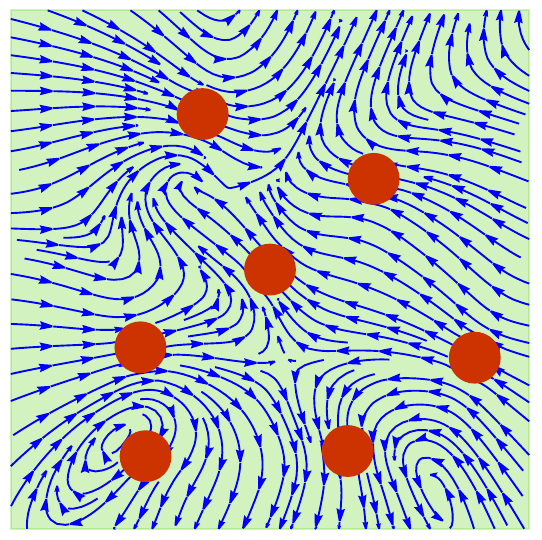}
  \end{tabular}
  \hspace*{\fill}
  \caption{Sketch of the monolayer within the fluid (particles in red;
    monolayer plane in green; fluid flow as blue streamlines): while
    the 3D flow is incompressible (left, ``laboratory perspective''),
    the particles experience only its projection onto the monolayer
    plane, which is 2D compressible (right, ``monolayer
    perpective'').}
  \label{fig:2dcomp}
\end{figure}

It is well established that the spatial dimensionality plays a
relevant role for the properties of physical systems both in and out
of equilibrium (see, e.g., \rcites{FoDi06,Ecke17}), a feature that has
motivated the research on the dependence of observables on the spatial
dimension. The dynamical properties of a colloidal monolayer immersed
in the bulk of a fluid provide an instance of experimental relevance:
while the hydrodynamic interactions (HI) mediated by the ambient fluid
propagate in three dimensions (3D), the colloidal particles are
constrained to move in a two--dimensional (2D) manifold due to some
trapping mechanism, e.g., wetting forces at a fluid interface, (see
the pioneering works \cite{Pier80,HuSc85,Onod85} and the works
\cite{LRW95,LCXZ14,MCDC21}, more relevant for our purpose), optical
traps \cite{BBSL02}, or phoretic forces \cite{WLBS22,SWBN25}.

The equilibrium structural properties of the monolayer are insensitive
to HI, but time--dependent observables, like the decay of equilibrium
fluctuations, or the properties of stationary but nonequilibrium
states are affected by HI. Therefore, transport coefficients have
become the most relevant observables for probing the influence of HI,
most notably the coefficients of self-diffusion (associated to the
mean square displacement of a tagged particle) and of collective
diffusion (describing the decay of perturbations in the colloidal
concentration), respectively, see, e.g., \rcite{Dhon96}.  In this
regard, the ``dimensional mismatch'' between the 3D-mediated HI and
the 2D monolayer\footnote{This configuration was dubbed ``partial
  confinement'' in \rcite{BDGH14} because the colloid is confined (to
  a plane), while the fluid is not.} has been shown
\cite{Banc99,NKPB02,BDGH14} to imply that the in-plane (2D) collective
diffusion of the monolayer becomes anomalous above a certain length
scale, meaning a breakdown of the monolayer Fick's law (colloidal
particle current proportional to the concentration gradient)
\cite{BDO15,BDO16}. This result is quite at odds with the fact that HI
induce only a finite renormalization of the value of the collective
diffusivity in the ``dimensionally plain'' scenarios of 3D colloid
within a 3D fluid \cite{Feld78,QWXP90,GeKl91,SePu96,VdC99,BHHN18}, and
2D colloid within a 2D fluid \cite{CLA04,FLVA04}. The predicted
anomaly has been observed experimentally \cite{LRW95,LCXZ14} and has
stimulated further theoretical study of the collective dynamics under
this kind of ``dimensional mismatch'': with time--dependent HI
\cite{Domi14}, in thick monolayers \cite{PPD17,BDO17}, in lipid
membranes \cite{PaDe18}, in nonplanar monolayers \cite{Domi18}, under
thermal fluctuations \cite{PBPX18}, close to a wall \cite{FCDD26}, in
monolayers of self-propelled particles \cite{SBSN24,YZZG26}.

The failure of Fick's law exhibits universal features that are
insensitive to the microscopic details of the colloid like
particle--particle forces or particle's shape and structure. Those
features can be traced back exclusively to the peculiar effect of HI
in this ``dimensionally mismatched'' configuration. First, while the
3D flow is incompressible, the colloidal particles only experience its
projection on the monolayer plane, which is a 2D compressible flow,
see \fref{fig:2dcomp}. Second, the ensuing HI has a long range that,
under the 2D compressibility feature, leads to a system--wide
interaction. The combination of these two properties leads to the
emergence of the anomalous collective diffusion, which can thus be
interpreted as a signature of a \textit{``compressible'', long--range
  hydrodynamic interaction}.

In this work, we extend the analytical study to the case of a binary
mixture of particles in the configuration ``colloidal monolayer + bulk
ambient fluid'' discussed above. In \sref{sec:single} we revisit the
single--component monolayer considered by previous works. In view of
the universality already mentioned, one can resort to some simplifying
approximations that isolate the effect of the ``compressible'' HI on
the anomaly of collective diffusion without loss of generality: the HI
is described faithfully at the mean--field level, and any interaction
via conservative forces is neglected (ideal--gas approximation). The
binary mixture is addressed in \sref{sec:binary} by applying these
same approximations. We conclude in \sref{sec:discussion} with a
discussion of results and future perspectives.

\section{Single--component monolayer}
\label{sec:single}

Consider a collection of identical colloidal particles constrained to
a monolayer immersed in a Newtonian fluid, characterized by a dynamic
viscosity $\eta$. Under the assumptions of ideality and overdamped
particle motion, and in the mean--field approximation for HI, the
Smoluchowski equation for the N--body distribution of particle
positions can be reduced to the following equation for the colloid
concentration, expressed through the particle number areal density
$\dens(\br,t)$ at a position $\br=(x,y)$ of the monolayer (and
$\nabla=(\partial_x, \partial_y)$ is the corresponding 2D nabla
operator) \cite{BDGH14,BDO15}:
\begin{subequations}
  \label{eq:dynamics}
\begin{equation}
  \label{eq:cont}
  \frac{\partial \dens}{\partial t} = - \nabla \cdot \bj ,
  \qquad
  \bj = - D \nabla \dens + \dens \bu .
\end{equation}
The particle current density $\bj$ has a contribution modeled by
Fick's law in terms of a (long--time collective) diffusion
coefficient\footnote{The generic dependence of $D$ on $n$ is neglected
  in the ideal--gas approximation. But even in the dilute limit, the
  value of $D$ is renormalized away from its single--particle behavior
  due to the interplay between hydrodynamic interactions and thermal
  fluctuations \cite{DFV14,PBPX18}.} $D$, and an advective piece due
to the in-plane hydrodynamic Stokes flow $\bu(\br)$ induced by the
Brownian diffusion current in the ideal--gas approximation:
\begin{equation}
  \label{eq:flow}
  \bu(\br) = \int d^2\br'\; \left[ - k_B T \, \nabla'\dens(\br') \right]
  \cdot \mathsf{O}(\br-\br') ,
\end{equation}
where the Oseen tensor (the Green function of the 3D Stokes equations
in an unbounded fluid) is given as
\begin{equation}
  \label{eq:oseen}
  \mathsf{O}(\br) = \frac{1}{8\pi\eta r} \left( \mathsf{I} +
    \frac{\br\br}{r^2} \right) 
\end{equation}
in dyadic notation (here $\mathsf{I}$ is the second-rank identity
tensor in 3D; however, since this tensor shows up contracted only with
a 2D field, one can interpret $\mathsf{I}$ also as the 2D identity
tensor as well).
\end{subequations}

\begin{figure}[t]
  \hfill
  \begin{tabular}[c]{ccc}
    \begin{tabular}[c]{c}
      \includegraphics[width=.11\textwidth]{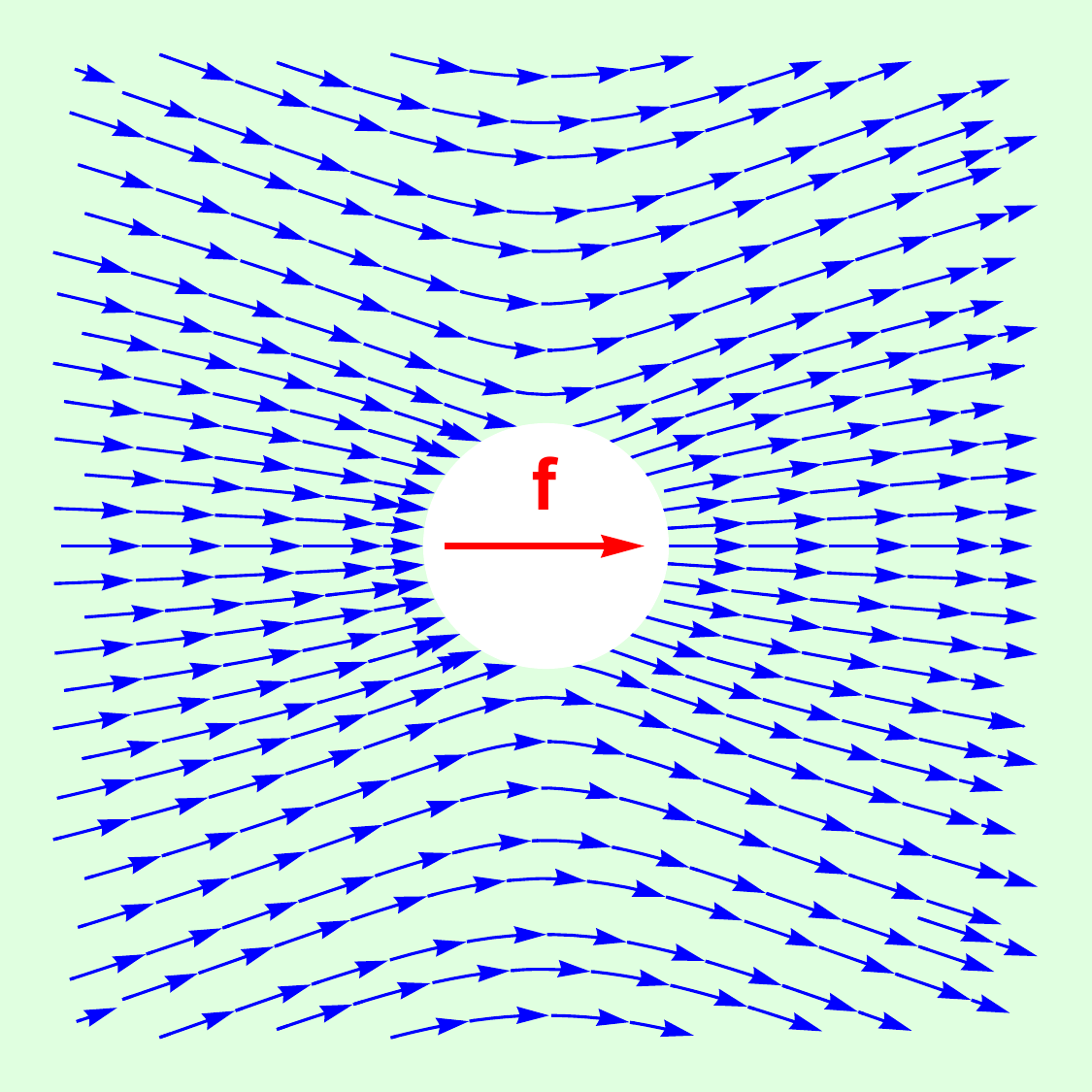}
    \end{tabular}
    & 
      \begin{tabular}[c]{c}
        \includegraphics[width=.24\textwidth]{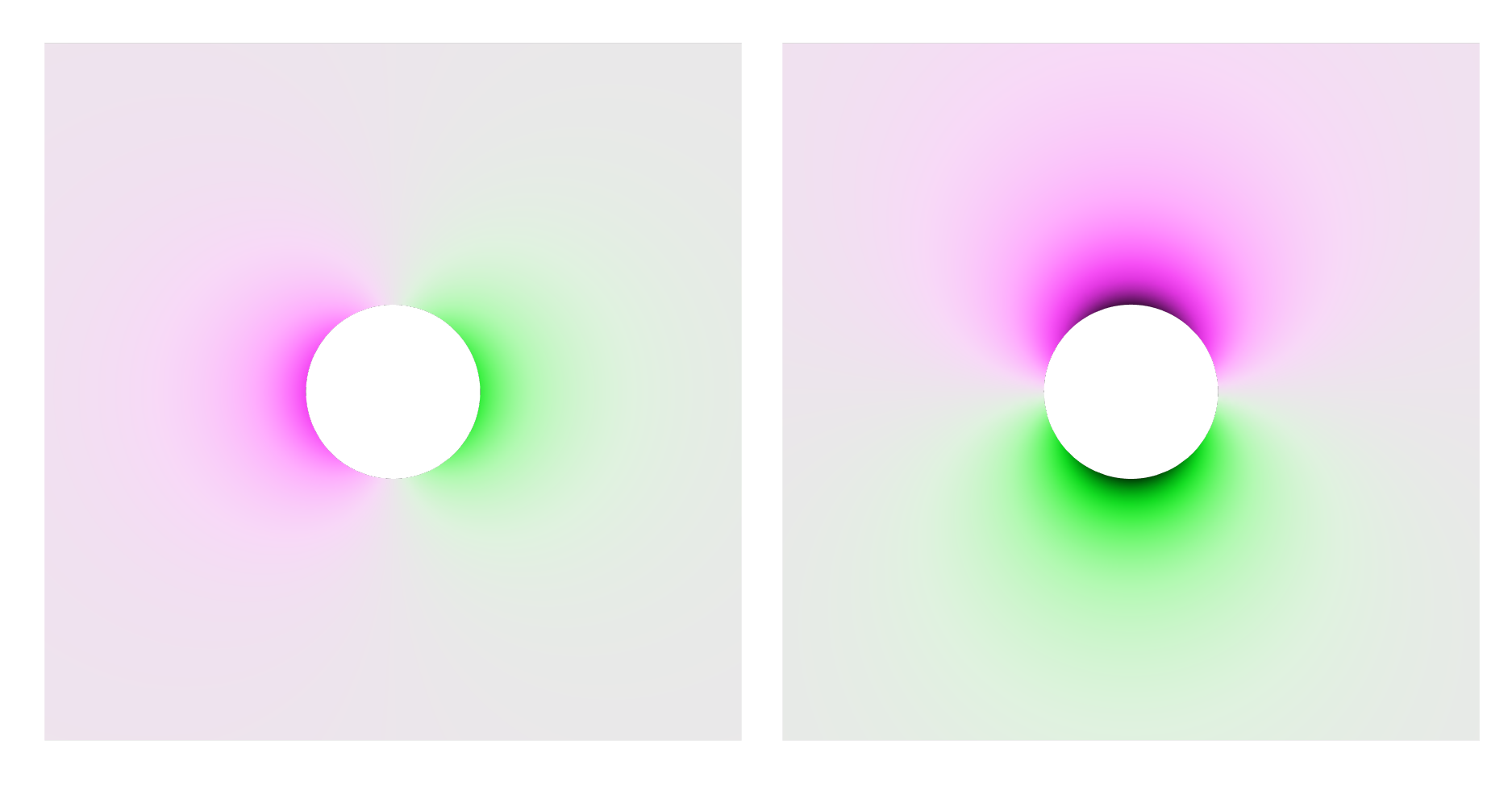}
      \end{tabular}
    &
      \begin{tabular}[c]{c}
        \includegraphics[width=.02\textwidth]{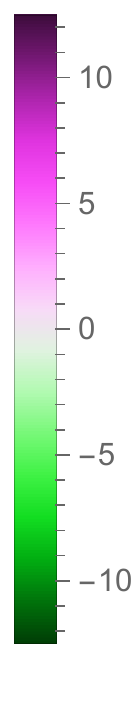}
      \end{tabular}
  \end{tabular}
  \hspace*{\fill}
  \caption{(\textit{Left}) Plot of the in-plane ($z=0$) streamlines of
    the flow field
    $\bu_\mathrm{oseen}(x,y)=\mathbf{f}\cdot\mathsf{O}(x,y)$ induced
    by an in-plane Stokeslet $\mathbf{f}$ applied at
    $(x,y)=(0,0)$. (\textit{Center}) The in-plane 2D divergence,
    $\nabla\cdot\bu_\mathrm{oseen}(x,y)$. (\textit{Right}) The
    in-plane vorticity, which has only a component normal to the
    monolayer plane,
    $\be_z\cdot\nabla\times\bu_\mathrm{oseen}(x,y)$. To facilitate the
    visualization, the region near the coordinate origin, where the
    fields diverge, has been omitted. The legend bar displays values
    in arbitrary units.}
  \label{fig:oseen}
\end{figure}

\begin{figure}[t]
  \includegraphics[width=.9\columnwidth]{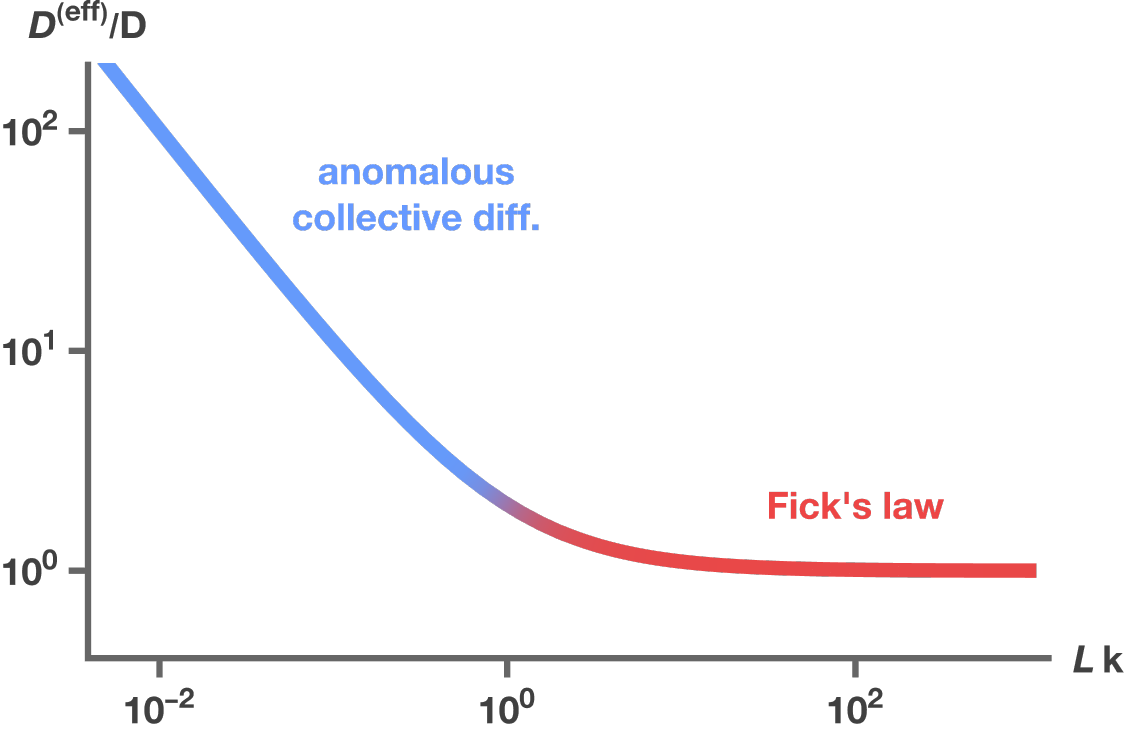}
  \caption{The effective collective diffusivity given by \eq{eq:Deff}
    as a function of the wave number, showing the crossover at
    $k^{-1}\sim L$ from Fick's law 
    to anomalous behavior.}
  \label{fig:Dsingle}
\end{figure}

This problem is qualitatively different from the case of a colloid in
bulk: although the flow stems from the \emph{incompressible} Stokes
equation in 3D, the tensor $\mathsf{O}(\br)$ describes a
\emph{compressible} 2D flow in the $z=0$ plane,
\begin{equation}
  \label{eq:divOseen}
  \nabla\cdot\mathsf{O}(\br) = - \frac{\br}{8\pi\eta r^3}
  = \nabla \left( \frac{1}{8\pi\eta r} \right) ,
\end{equation}
and, correspondingly, it is $\nabla\cdot\bu\neq 0$, see
\fref{fig:oseen} --- notice that \eq{eq:oseen} is not the Oseen tensor
associated to the 2D Stokes equation. Furthermore, the last equality
in \eq{eq:divOseen} shows that the collective in-plane flow
field~(\ref{eq:flow}) driven by the number density gradient is
actually a 2D potential flow,
\begin{subequations}
 \label{eq:coulomb} 
 \begin{equation}
   \label{eq:flow2}
   \bu(\br) = - k_B T 
   \int d^2\br'\; \dens(\br') \nabla\cdot\mathsf{O}(\br-\br')
   = - \nabla\Psi(\br) ,
 \end{equation}
 after integrating by parts and defining the scalar potential
 \begin{equation}
   \label{eq:Psi}
   \Psi(\br) := \frac{k_B T}{8\pi\eta} \int d^2 \br' \;
   \frac{\dens(\br')}{|\br-\br'|} .
 \end{equation}
\end{subequations}
In this disguise, the advective contribution to the particle current
looks formally like 3D Coulombic repulsion, which becomes responsible for the enhanced
(actually divergent) diffusivity. Indeed, one can address the decay of
deviations $\dn(\br)=\dens(\br)-\dens^{(0)}$ from an (equilibrium)
homogeneous state $\dens^{(0)}$: upon linearizing \eqs{eq:dynamics} for
$|\dn|\ll\dens^{(0)}$, one gets
\begin{equation}
  \label{eq:lin}
  \frac{\partial \dn}{\partial t} \approx D \nabla^2 \dn
  - \dens^{(0)} \nabla\cdot\bu .
\end{equation}
By introducing the 2D Fourier transform,
\begin{equation}
  \label{eq:Fourier}
  \hat{\dens}(\bk,t) := \int d^2\br\; \mathrm{e}^{-i\bk\cdot\br} \, \dn(\br,t) ,
\end{equation}
the linearized dynamics can be formulated as
\begin{equation}
  \label{eq:linFourier}
  \frac{\partial \hat{\dens}}{\partial t} = - k^2 D^\mathrm{(eff)} (k)
  \, \hat{\dens} , 
\end{equation}
in terms of an effective, wave-number--dependent coefficient of
collective diffusion,
\begin{equation}
  \label{eq:Deff}
  \frac{D^\mathrm{(eff)}(k)}{D} := 1 + \pec ,
  \quad
  \pec := \frac{1}{L k}  ,
  \quad
  L := \frac{4\eta D}{\dens^{(0)}\, k_B T} .
\end{equation}
Here, $L$ is a characteristic \textit{hydrodynamic length}
\cite{BDGH14}, that allows one to define $\pec$ as the
wave-number--dependent Peclet number that measures the importance of
advection relative to Brownian diffusion at the length scale $k^{-1}$.
The divergence at $k=0$ signals the breakdown of Fick's law, in the
form of an anomalous collective diffusion which sets in on spatial
scales above the length $L$ (i.e., for large Peclet number), see
\fref{fig:Dsingle}. The specific form of this divergence is universal
in the sense that it is not altered \cite{BDO15,BDO16} by introducing
corrections either to the ideal--gas approximation or to the
mean--field approximation for the HI, both of which show up in a
possible $n$--dependence of the finite ``bare'' diffusivity $D$. The $k^{-1}$
divergence has been confirmed by experimental measurements
\cite{LRW95,LCXZ14}; this differs significantly from the effect of HI
interactions in ``dimensionally plain'' cases (3D colloid within 3D
fluid, or 2D colloid within 2D fluid), where the incompressibility
constraint on the flow leads to HI--induced finite corrections of the
collective ``bare'' diffusivity $D$
\cite{Feld78,QWXP90,GeKl91,SePu96,VdC99,BHHN18,CLA04,FLVA04}.

Finally, we note that a variational reformulation of the model is
available. One can define the functional
\begin{subequations}
  \label{eq:functionalFormulation}
\begin{equation}
  \label{eq:freeF}
  \mathcal{F}[n] := \int d^2\br\; \left\{
    D \, \dens(\br) \left[ \ln \frac{\dens(\br)}{\dens^{(0)}} - 1 \right]
    + \frac{1}{2} \dens(\br) \Psi(\br)
  \right\} ,
\end{equation}
in terms of the field~(\ref{eq:Psi}), and express the current density
in \eq{eq:cont} as
\begin{equation}
  \label{eq:newj}
  \bj(\br) = - \dens(\br) \,
  \nabla \, \frac{\delta \mathcal{F}[\dens]}{\delta \dens(\br)} .
\end{equation}
\end{subequations}
The functional $\mathcal{F}[\dens]$ evolves monotonically in time
according to the dynamics described by \eqs{eq:dynamics}:
\begin{eqnarray}
  \label{eq:dynF}
  \frac{d \mathcal{F}}{d t} 
  & =
  & \int d^2\br\; \frac{\delta \mathcal{F}[\dens]}{\delta \dens(\br)}
    \, \frac{\partial \dens(\br)}{\partial t}
    \nonumber
  \\
  & =
  & - \int d^2\br\;  n(\br) \,
    \left| \nabla \, \frac{\delta \mathcal{F}[n]}{\delta n(\br)}
    \right|^2 \leq 0 .
\end{eqnarray}
Since $\mathcal{F}$ is bounded from below ($\Psi\geq 0$,
$x (\ln x-1) \geq -1$), it is actually a Lyapunov functional and the
equilibrium state, $n(\br,t)=\dens^{(0)}$, is an attractor of the
dynamics that minimizes the functional $\mathcal{F}[n]$.

One can recognize in this functional the free energy of a gas of
equal--signed point charges in the mean--field approximation: there is
an ideal--gas contribution while the field $\Psi$, that represents the
HI, resembles an effective Coulombic repulsion. The functional
$\mathcal{F}[n]$ could be dubbed a pseudo free energy, but this
approach is not very useful because the existence of $\mathcal{F}$ is
a very specific feature of the ideal--gas approximation: as soon as
one accounts for direct interactions, this is no longer the case as a
consequence of the nonreciprocal nature of HI, see App.~\ref{app:noF}.

\section{Binary--component monolayer}
\label{sec:binary}

The model can be extended to the case of a monolayer consisting of a
mixture of two kinds of different particles, which we term ``big'' (or
``slow'') and ``small'' (or ``fast'') according to their respective
diffusivities, $\Db$ and $\Ds\geq \Db$. The dynamic evolution of the
corresponding number densities, $\nb(\br,t)$ and $\ns(\br,t)$, is described
by the equations
\begin{subequations}
  \label{eq:dynamicsbin}
\begin{eqnarray}
  \frac{\partial \nb}{\partial t} = - \nabla \cdot \bj_\mathrm{b} ,
  & 
  & \bj_\mathrm{b} = - \Db \nabla \nb + \nb \bu ,
    \label{eq:contbig}
  \\
  \nonumber \\
  \frac{\partial \ns}{\partial t} = - \nabla \cdot \bj_\mathrm{s} ,
  & 
  & \bj_\mathrm{s} = - \Ds \nabla \ns + \ns \bu ,
    \qquad
    \label{eq:contsmall}
\end{eqnarray}
with an in-plane flow generated by the distribution of both kinds of
particles,
  \begin{equation}
    \label{eq:flow2bin}
    \bu(\br) = - k_B T \int d^2\br'\; \nabla' \left[
      \nb(\br') + \ns(\br')
    \right]
    \cdot \mathsf{O}(\br-\br') ,
  \end{equation}
  in terms of the Oseen tensor~(\ref{eq:oseen}).
\end{subequations}
The distributions of each type of particle in the mixture are
therefore coupled
through the drag by the common flow $\bu(\br)$.
Like in the single--component case, there also exists an alternative
formulation based on a pseudo free energy that accounts for the
coupling by HI:
\begin{subequations}
\begin{align}
  \label{eq:freeFbin}
  \mathcal{F}[\nb,\ns] := \int d^2\br\;
  & \left\{
    \Db \nb(\br) \left[ \ln
    \frac{\nb(\br)}{\nb^{(0)}} - 1 \right]
    \right.
    \nonumber \\
  & + \Ds \ns(\br) \left[ \ln
    \frac{\ns(\br)}{\ns^{(0)}} - 1 \right]
    \nonumber \\
  & \left. + \frac{1}{2} \left[ \nb(\br) + \ns(\br)
    \right] \Psi(\br)
    \right\} ,
    \qquad
\end{align}
\begin{equation}
  \label{eq:Psibin}
  \Psi(\br) := \frac{k_B T}{8\pi\eta} \int d^2 \br' \;
  \frac{\nb(\br') + \ns(\br')}{|\br-\br'|},
\end{equation}
\end{subequations}
 so that one can write 
\begin{equation}
  \label{eq:newjbin}
  \bj_\mathrm{\alpha}(\br) = - \dens_\mathrm{\alpha}(\br) \,
  \nabla \, \frac{\delta \mathcal{F}[\nb, \ns]}{\delta \dens_\mathrm{\alpha}(\br)} ,
  \qquad
  \alpha=\mathrm{b,s}.
\end{equation}
The functional $\mathcal{F}$ is again a Lyapunov functional of the
dynamics described by \eqs{eq:dynamicsbin}.

The evolution of the small deviations from a homogeneous mixture can
be derived by linearizing the dynamic equations in the deviations
$\dn_\mathrm{\alpha} = \dens_\mathrm{\alpha} -
\dens_\mathrm{\alpha}^{(0)}$ away from a homogeneous state
characterized by the number densities $\ns^{(0)}$ and $\nb^{(0)}$. The
resulting expressions can be expressed conveniently in terms of
dimensionless parameters: the ratios of diffusivities and average
densities, respectively,
\begin{equation}
  \label{eq:ratios}
  \delta := \frac{\Ds}{\Db} \geq 1,
  \qquad
  \nu := \frac{\ns^{(0)}}{\nb^{(0)}} ,
\end{equation}
and the Peclet numbers associated to a given wave number $k$ for big
and small particles, respectively, as if each kind were unmixed, see
\eq{eq:Deff}:
\begin{subequations}
  \label{eq:peclet}
\begin{equation}
  \label{eq:Peb}
  \Peb := \frac{1}{\Lb k} ,
  \qquad
  \Lb := \frac{4\,\eta\, \Db}{\nb^{(0)}\, k_B T} ,
\end{equation}
and 
\begin{equation}
  \label{eq:Pes}
  \Pes := \frac{1}{\Ls k} ,
  \qquad
  \Ls := \frac{4\,\eta\, \Ds}{\ns^{(0)}\, k_B T} .
\end{equation}
These parameters are related as
\begin{equation}
  \label{eq:ratioPe}
  \frac{\Pes}{\Peb} = \frac{\Lb}{\Ls}
  = \frac{\nu}{\delta} , 
\end{equation}
so that there are only three independent parameters pertaining to the
linear dynamics, which we conveniently choose to be $\delta$, $\nu$,
and $\Peb$, see \eqs{eq:eigen}.
\end{subequations}

Therefore, upon introducing the Fourier transforms,
\begin{equation}
  \label{eq:Fourierbin}
  \nhvar{\alpha}(\bk, t) := \int d^2\br\;
  \mathrm{e}^{-i\bk\cdot\br} \, \dn_\mathrm{\alpha}(\br,t) ,
  \qquad
  \alpha=\mathrm{b,s},
\end{equation}
one gets
\begin{subequations}
  \label{eq:linEq}
\begin{equation}
  \label{eq:linearMixture}
  \frac{\partial}{\partial t} \left(
    \begin{array}[c]{c}
      \nhb
      \\
      \\
      \nhs
    \end{array}
  \right)
  = -k^2 \Db \mathsf{M}
  \left(
    \begin{array}[c]{c}
      \nhb
      \\
      \\
      \nhs
    \end{array}
  \right) ,
\end{equation}
with the wave-number--dependent matrix\footnote{For notational
  simplicity, we omit an explicit indication of the dependence on the
  wave number $k$ (via the parameter $\Peb$) for the matrix and for
  its eigenvectors and eigenvalues.}
\begin{equation}
  \label{eq:matr}
  \mathsf{M} = \left(
    \begin{array}[c]{cc}
      \displaystyle 1
      & 0
      \\
      \\
      0
      & \displaystyle \delta 
    \end{array}
  \right)
  + \Peb \left(
    \begin{array}[c]{cc}
      \displaystyle 1
      & \displaystyle 1
      \\
      \\
      \displaystyle \nu
      & \displaystyle \nu
    \end{array}
  \right),
\end{equation}
\end{subequations}
that has been split naturally into the piece stemming from ideal--gas
diffusion and the contribution by HI, compare with \eq{eq:Deff}. The
solution to \eq{eq:linearMixture} can be written as
\begin{equation}
  \label{eq:gensol}
  \left(
    \begin{array}[c]{c}
      \nhb
      \\
      \\
      \nhs
    \end{array}
  \right)
  = C_+ \, \vec{\xi}_+ \, \mathrm{e}^{-t  k^2 \Db \Lambda_+} 
  + C_- \, \vec{\xi}_- \, \mathrm{e}^{- t  k^2 \Db \Lambda_-},
\end{equation}
in terms of two integration constants $C_\pm$ and the eigenvalues and
corresponding eigenvectors of $\mathsf{M}$:
\begin{subequations}
  \label{eq:eigen}
\begin{equation}
  \Lambda_\pm := \frac{1+\delta+ (1+\nu) \Peb}{2} \pm \Delta ,
\end{equation}
\begin{equation}
  \vec{\xi}_\pm := \left(
    \begin{array}[c]{c}
      \displaystyle \frac{1-\delta+ (1-\nu) \Peb}{2} \pm \Delta
      \\
      \\
      \nu\, \Peb
    \end{array}
  \right) ,
\end{equation}
\begin{equation}
  \Delta := \sqrt{\left[\frac{1-\delta+(1-\nu) \Peb}{2}\right]^2 + \nu\, \Peb^2} .
\end{equation}
\end{subequations}
In order to gain insight into this solution, we consider two different
physically meaningful, limiting cases.

\subsection{Identical particles}
\label{sec:limit-identical}

When the particles are dynamically identical, it holds $\delta=1$ and
the only difference is a mechanically irrelevant particle labelling,
which is conventionally termed ``color''. This system configuration is
usually applied to single out and address the dynamics of a particle;
specifically, in the limit that one of the particle types is very
dilute, e.g., $\nu \to 0$, it becomes a tracer and collective and
self--diffusion become indistinguishable for it.  This approach was
applied in \rcite{PBPX18} to study the role of HI in the
``dimensionally mismatched'' monolayer configuration.

Therefore, setting $\delta=1$ and taking the limit $\nu\to 0$ at fixed
$\Peb$ in the expressions~(\ref{eq:eigen}), they become\footnote{Any
  piece that can be factored out from the eigenvectors would be
  trivially incorporated in the corresponding integration constant
  $C_\pm$ that appears in the general solution~(\ref{eq:gensol}). This
  explains our use of the equivalence symbol ($\equiv$), rather than
  the equality, in order to emphasize the direction of the
  eigenvectors, which is the only relevant feature.}
\begin{subequations}
  \label{eq:tracereigen}
\begin{equation}
  \label{eq:tracerplus}
  \Lambda_+=1+\Peb,
  \qquad
  \vec{\xi}_+ \equiv \left(
    \begin{array}[c]{c}
      \displaystyle 1
      \\
      0
    \end{array}
  \right) ,
\end{equation}
\begin{equation}
  \label{eq:tracerminus}
  \Lambda_-=1 ,
  \qquad
  \vec{\xi}_- \equiv \left(
    \begin{array}[c]{c}
      \displaystyle -1
      \\
      1
    \end{array}
  \right) .
\end{equation}
\end{subequations}
The mode $\vec{\xi}_+$ describes the evolution of the big particle
distribution in the absence of tracers and, unsurprisingly, agrees
completely with the single--component scenario --- compare $\Lambda_+$
with \eq{eq:Deff}. The mode $\vec{\xi}_-$ entails the evolution of the
tracer (a ``small'' particle),
\begin{equation}
  \label{eq:nsmallcolor}
  \nhs (\bk,t) \sim \mathrm{e}^{- t  k^2 \Db} .
\end{equation}
This result shows that the anomalous collective diffusion induced by
HI is a truly collective effect associated to a local redistribution
of particles, not to an anomalous diffusion of the single particles,
whose self-diffusion remains finite. This conclusion was reported in
\rcite{PBPX18} from numerical simulations; the coefficient of
self-diffusion does get a finite correction from HI, but only when
going beyond the mean--field approximation \cite{PBPX18,Domi25}.

\subsection{Strongly dissimilar particles}
\label{sec:limit-very-different}

In the opposite case of particles with very different diffusivities
($\delta\gg 1$), one has to consider a distinguished limit if the HI
are to play a relevant role: in view of the structure~(\ref{eq:matr})
of the matrix for the linear dynamics, one must also allow for a
distribution of small particles much less dilute than the big ones,
and take the double limit
\begin{equation}
  \label{eq:distlim}
  \left.
    \begin{array}[c]{c}
      \delta\to +\infty
      \\
      \nu \to +\infty
    \end{array}
  \right\}
  \quad
  \mathrm{but}\; \Peb \; \mathrm{and} \;
  \frac{\Pes}{\Peb} = \frac{\nu}{\delta} \;\mathrm{finite}.
\end{equation}
In this case, the expressions~(\ref{eq:eigen}) simplify to 
\begin{subequations}
  \label{eq:surfactanteigen}
\begin{equation}
  \label{eq:surfactanctplus}
  \Lambda_+=\delta \left( 1 + \Pes \right) ,
  \qquad
  \vec{\xi}_+ \equiv \left(
    \begin{array}[c]{c}
      \displaystyle 0
      \\
      1
    \end{array}
  \right) ,
\end{equation}
\begin{equation}
  \label{eq:surfactanctminus}
  \Lambda_- = 1 + \frac{\Peb}{1  + \Pes}  ,
  \qquad
  \vec{\xi}_- \equiv \left(
    \begin{array}[c]{c}
      \displaystyle 1 + 1/\Pes
      \\
      -1
    \end{array}
  \right) ,
\end{equation}
\end{subequations}
and, correspondingly, the particle distributions evolve as (see
\eq{eq:gensol})
\begin{subequations}
  \label{eq:solutionSurf}
\begin{equation}
  \label{eq:nsSurf}
  \nhs(\bk,t) = \nhs^\mathrm{(adia)}(\bk,t)+ \nhs^\mathrm{(pure)} (\bk,t) ,
\end{equation}
\begin{equation}
  \label{eq:nsSurfAdia}
  \nhs^\mathrm{(adia)}(\bk,t) := -\frac{\nhb(\bk,t)}{1+1/\Pes(k)} ,
\end{equation}
\begin{equation}
  \label{eq:nsSurfPure}
  \nhs^\mathrm{(pure)}(\bk,t) \propto \mathrm{e}^{- t  k^2 D_\mathrm{s}^\mathrm{(pure)}(k)} , 
\end{equation}
\begin{equation}
  \label{eq:nbSurf}
  \nhb(\bk,t) \propto \mathrm{e}^{-t k^2 D_\mathrm{b}^\mathrm{(eff)}(k)} ,
\end{equation}
\end{subequations}
after defining the wave-number--dependent diffusivities
\begin{subequations}
\begin{equation}
  \label{eq:Dplus}
  D_\mathrm{s}^\mathrm{(pure)}(k) := \Db \Lambda_+(k)
  = \Ds \left( 1 + \frac{1}{\Ls k} \right) ,
\end{equation}
and
\begin{equation}
  \label{eq:Dminus}
  D_\mathrm{b}^\mathrm{(eff)}(k) := \Db \Lambda_ -(k)
  = \Db
  \left[
    1 + \frac{1}{(\nu/\delta) + \Lb k}
  \right].
\end{equation}
\end{subequations}
Therefore, the mode $\vec{\xi}_+$, which corresponds to the
$\nhs^\mathrm{(pure)}$ term, describes the \emph{pure} (i.e., unmixed)
evolution of the small particle distribution, with a time scale set
by the diffusivity $D_\mathrm{s}^\mathrm{(pure)}$, 
compare with \eq{eq:Deff}. The mode $\vec{\xi}_-$, on the contrary,
describes the joint evolution of big and small particles coupled by
HI: the dynamic evolution of the big particles evolution is
characterized by the effective collective diffusivity
$D_\mathrm{b}^\mathrm{(eff)}$, whereas the $\nhs^\mathrm{(adia)}$ term
represents the \emph{adia}batic enslaving of the small particle
distribution to that of the big particles: indeed, it turns out to be
the stationary solution of the dynamical equation for $\nhs(\bk,t)$,
\begin{equation}
  \label{eq:nsDyn}
  \frac{\partial \nhs}{\partial t}
  = - k^2 D_\mathrm{s}^\mathrm{(pure)} (k) \left[
    \nhs + \frac{\Pes(k)\, \delta}{\Lambda_+(k)} \, \nhb
  \right] ,
\end{equation}
see \eqs{eq:linEq}, when $\nhb(\bk,t)$ can be assumed to be
approximately constant in time (or, more precisely, when its evolution
on the time scale set by $D_\mathrm{s}^\mathrm{(pure)}(k)$ is slow).

\begin{figure}[t]
  \includegraphics[width=.9\columnwidth]{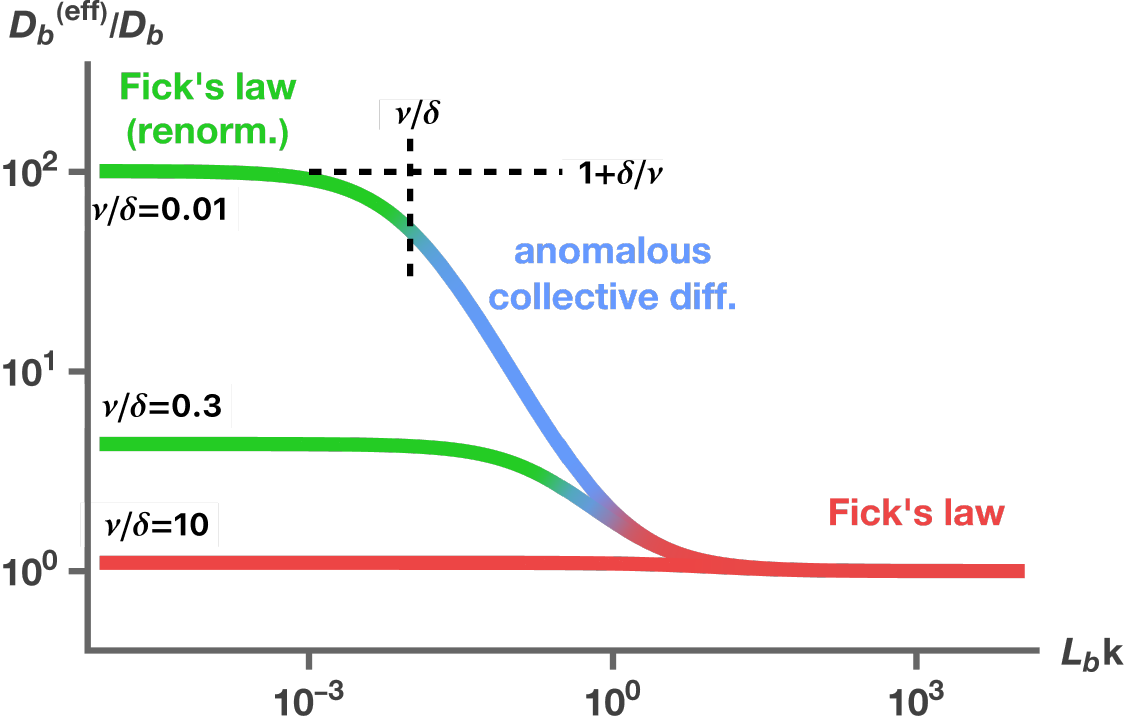}
  \caption{The effective collective diffusivity of the big particles,
    given by \eq{eq:Dminus}, as a function of the wave number for
    different values of the ratio $\nu/\delta$, see
    \eq{eq:ratioPe}. In addition to Fick's law and anomalous behavior,
    as in the single component case (compare with \fref{fig:Dsingle}),
    one identifies a regime of renormalized (enhanced) Fick's law. See also Table
    \ref{tab:regimes}.}
  \label{fig:Deff}
\end{figure}

\begingroup
\squeezetable
  \begin{table*}[t]
    \centering

    \begin{ruledtabular}
    \begin{tabular}[t]{c||c|c|c|c||}
      \multirow{3}{*}{
      \bf \begin{tabular}[c]{c}
            dynamics of
            \\
            small particles
      \end{tabular}
      }
      & \multicolumn{2}{c|}{$\Pes \ll 1 \quad (k^{-1} \ll \Ls)$}
      & \multicolumn{2}{c||}{$1 \ll \Pes \quad (\Ls \ll k^{-1})$}
      \\
      \cline{2-5}
      & \multicolumn{2}{c|}{
          pure Fick's law
          as if unmixed
      }
      & \multicolumn{2}{c||}{
        $\displaystyle
        \textrm{HI--induced}
        \left\{
        \begin{array}[c]{c}
          \textrm{anomalous collective diff.}
          \\
          \textrm{ + adiabatic enslaving}
        \end{array}
      \right.
      $
      }
      \\
      \hline\hline
      \multirow{3}{*}{
      \bf \begin{tabular}[c]{c}
        dynamics of
        \\
        big particles
      \end{tabular}
      }
      &  
        \colorbox{myred!90}{
        \textcolor{white}{
        \begin{tabular}[c]{c}
          pure Fick's law
          \\
          $\displaystyle \phantom{\Bigg|}
          D_\mathrm{b}^\mathrm{(eff)}(k) \approx \Db \phantom{\Bigg|}$
       \end{tabular}
      }
      }
      & \colorbox{myblue!90}{
        \textcolor{white}{
        \begin{tabular}[c]{c}
          anomalous collective diff. 
          \\
          $\displaystyle \phantom{\Bigg|} D_\mathrm{b}^\mathrm{(eff)}(k) \approx
          \frac{\Db}{\Lb k}$
        \end{tabular}
      }
      }
      & \colorbox{myred!90}{
        \textcolor{white}{
        \begin{tabular}[c]{c}
          pure Fick's law
          \\
          $\displaystyle \phantom{\Bigg|}
          D_\mathrm{b}^\mathrm{(eff)}(k) \approx \Db \phantom{\Bigg|}$
        \end{tabular}
      }
      }
      & \colorbox{mygreen!90}{
        \textcolor{white}{
        \begin{tabular}[c]{c}
           renormalized Fick's law
           \\
           $\displaystyle \phantom{\Bigg|} D_\mathrm{b}^\mathrm{(eff)}(k) \approx
           \Db \frac{\Ls}{\Lb} \gg \Db \phantom{\Bigg|}$
        \end{tabular}
      }
      }
      \\
      \cline{2-5}
      & $\Peb \ll 1$
      & $1\ll \Peb$
      & $1 \ll \Peb \ll \Pes$
      & $1 \ll \Pes \ll \Peb$
      \\
      & $(k^{-1} \ll \Lb)$
      & $(\Lb \ll k^{-1})$
      & $(\Ls \ll \Lb \ll k^{-1})$
      & $(\Lb \ll \Ls \ll k^{-1})$
      \\
    \end{tabular}
  \end{ruledtabular}
    
  \caption{The different dynamic regimes for the case of strongly
    dissimilar particles ($\delta, \nu \gg 1$), see also
    Figs.~\ref{fig:Deff} and \ref{fig:Peplane}. The definitions of the
    parameters are given by \eqs{eq:ratios} and (\ref{eq:peclet}) and
    the probing length scale $k^{-1}$ is set by the wave number.}
    
\label{tab:regimes}
\end{table*}
\endgroup

The effective diffusivity of the big particles is shown in
\fref{fig:Deff}, which allows one to identify different dynamic
regimes depending on the values of the parameters, as summarized in
Table \ref{tab:regimes}. This can be visualized easily in the plane
spanned by the parameters $\Peb$ and $\Pes$, see
\fref{fig:Peplane}. In this diagram, the limit of no HI at all, which
is obtained formally by letting $\eta\to\infty$ (i.e., setting Oseen
tensor to zero) unconditionally in the model \eqs{eq:dynamicsbin},
corresponds to the origin, $\Peb=\Pes=0$, see \eqs{eq:peclet}. When HI
play a role, the dynamical regimes are grossly separated by ``fuzzy''
boundaries across which the transition in the dynamical behavior is
smooth.

When $\Pes \ll 1$ (that is, at small scales for the small particles),
the evolution of the two kinds of particles is effectively decoupled,
as if unmixed: the small particles neither notice the big ones
($\nhs^\mathrm{(adia)}\approx 0$ in \eq{eq:nsSurf}) nor experience HI
($D_\mathrm{s}^\mathrm{(pure)} \approx \Db$),
whereas the big particles evolve as in the pure case under the effect
of the HI among themselves,
\begin{equation}
  \label{eq:Dbpure}
  \frac{D_\mathrm{b}^\mathrm{(eff)}}{\Db} 
  \approx 1 + \Peb ,
\end{equation}
compare with \eq{eq:Deff}. Therefore, at sufficiently small scales,
the presence of small particles is irrelevant for the big ones, which,
depending on the value of $\Peb$, will appear to exhibit either normal
or anomalous collective diffusion.

A dynamic coupling mediated by HI only occurs in the opposite limit
$\Pes \gg 1$ (large scales for the small particles): the big particles
follow Fick's law (i.e., no anomalous collective diffusion), albeit
with a possibly renormalized diffusivity,
\begin{equation}
  \label{eq:DbFick}
  \frac{D_\mathrm{b}^\mathrm{(eff)}}{\Db} 
  \approx 1 + \frac{\Peb}{\Pes} .
\end{equation}
This becomes indistinguishable of the big particle ``bare''
diffusivity in the unmixed (pure) case when, additionally,
$\Peb\ll\Pes$. The relevant feature is that the adiabatic component of
the small particle distribution adjusts perfectly to the big particle
distribution, $\nhs^\mathrm{(adia)} \approx - \nhb$, and consequently
the role of $\nhb$ as source of flow in \eq{eq:flow2bin} becomes less
and less important when $\Pes\to \infty$. We emphasize, however, that
the 2D compressible in-plane flow remains nevertheless relevant, being
ultimately responsible for this renormalized diffusivity as a
consequence of the much faster, anomalous collective diffusion of the
small particles:
\begin{equation}
  \label{eq:Dbpureanom}
  \frac{D_\mathrm{b}^\mathrm{(eff)}}{D_\mathrm{s}^\mathrm{(pure)}}
  \approx \frac{\Db ( 1 + \Peb/\Pes)}{\Ds\, \Pes}
  = \frac{1}{\Pes} \left( \frac{1}{\delta} + \frac{1}{\nu} \right)
  \ll 1 .
\end{equation}

\section{Discussion}
\label{sec:discussion}

We have analyzed the collective dynamics of a binary colloidal
monolayer in the bulk of a fluid, setting the focus on the role of the
long--range hydrodynamic coupling between the mixture components.
Depending on the relative values of the system's parameters, we have
addressed different physically meaningful scenarios. In the limit of
identical particles, our result confirms the numerical conclusion of
\rcite{PBPX18} that the deviation from Fick's law is a truly
collective effect, in the sense that the anomalous decay of
concentration fluctuations succeeds although the single--particle
diffusion is not anomalous.

\begin{figure*}[t]
  \hfill
  \includegraphics[width=.9\columnwidth]{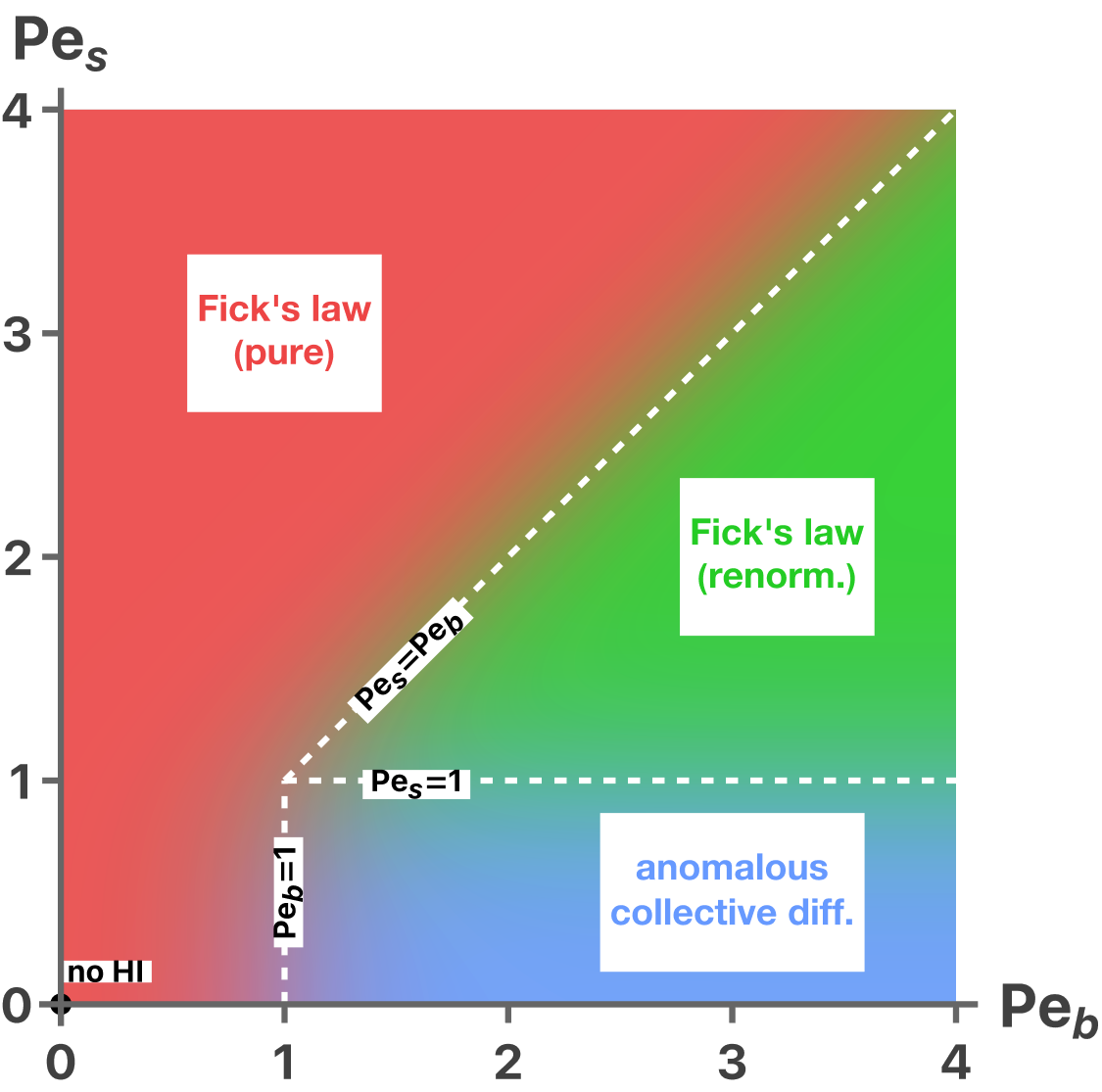}
  \hfill
  \includegraphics[width=.9\columnwidth]{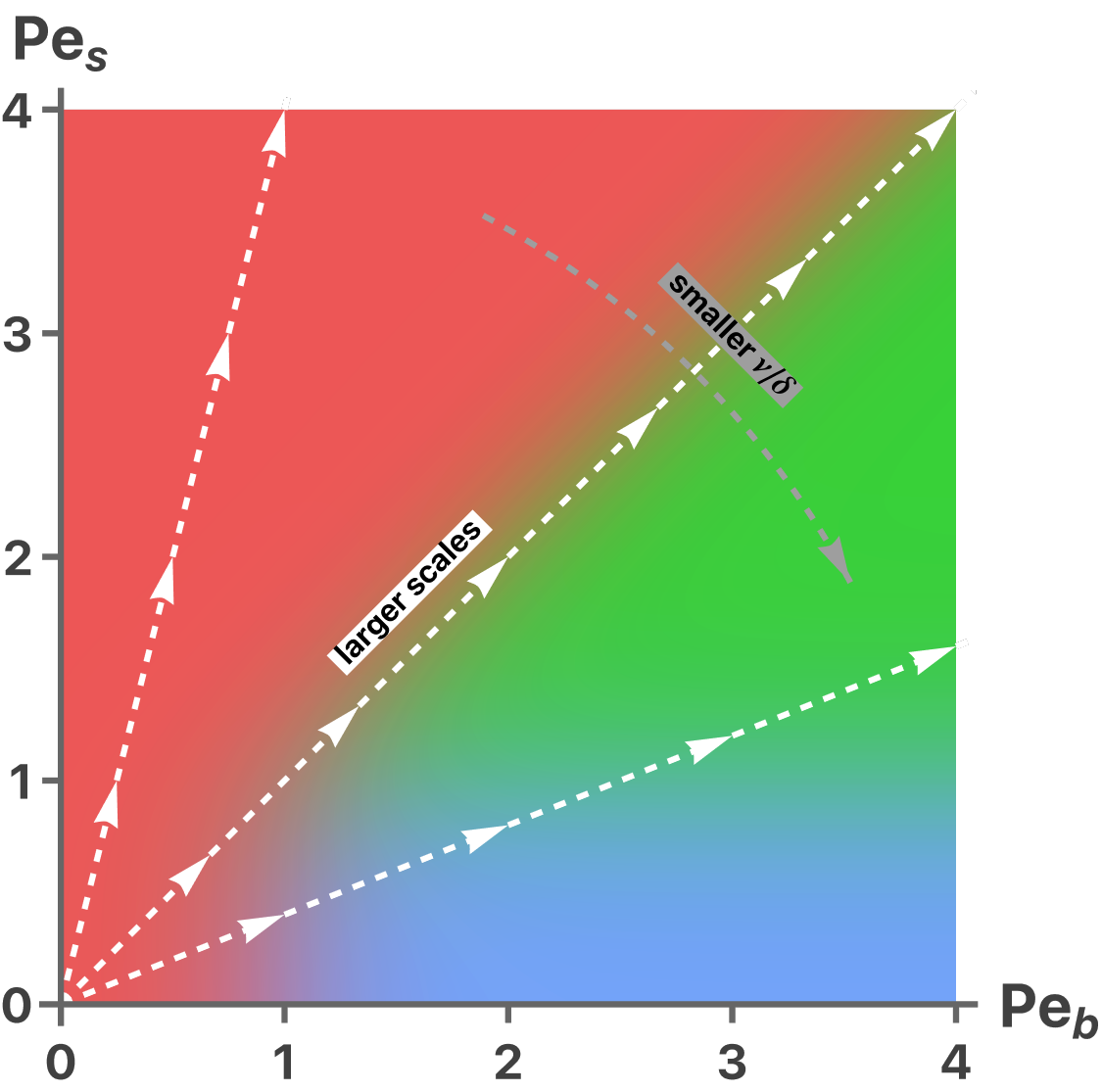}
  \hspace*{\fill}
  \caption{(Left) Parameter plane $(\Peb, \Pes)$, depicting the
    regimes for the effective dynamics of the big particles in a
    mixture of two strongly dissimilar kinds of particles
    ($\delta, \nu \gg 1$). (Right) Probing the parameter plane
    (compare with Table~\ref{tab:regimes}): increasing the length
    scale of the probe ($k\to 0$) at fixed material parameters is
    represented by a shift (white arrows) along a straight line that
    starts at the coordinate origin and whose slope is determined by
    the parameter $\nu/\delta=\Lb/\Ls$.}
  \label{fig:Peplane}
\end{figure*}

In the opposite limit of strongly dissimilar particles, the effective
dynamics of the big particles is summarized in \fref{fig:Peplane}, where
three different regimes are identified
as captured by Eqs.~(\ref{eq:Dbpure}) and
(\ref{eq:DbFick}). Well inside the red region, it is
$D_\mathrm{b}^\mathrm{(eff)}\approx \Db$, i.e., the big particles
appear to obey Fick's law as if unmixed and in the absence of HI. Well
inside the blue region it holds
$D_\mathrm{b}^\mathrm{(eff)}\approx \Db \Peb$, that is, the big
particles exhibit the anomalous collective diffusion due to HI
characteristic of the single--component case, compare with
\eq{eq:Deff}. And finally, deep in the green region, the big particles
obey Fick's law effectively, but they appear ``vested'': due to the
hydrodynamic coupling to the small particles, the effect of 2D
compressibility shows up as an enhanced, but finite collective
diffusivity, which becomes actually controlled
by the diffusivity of the small particles:
$D_\mathrm{b}^\mathrm{(eff)}\approx (\Peb/\Pes) \Db \approx
(\nb^{(0)}/\ns^{(0)})\Ds \gg \Db$.

Accordingly, in order to have a manifest effect of the HI on the
collective dynamics of the big particles, one necessarily needs
$\Peb \gtrsim 1$, just like in the pure limit case (no mixture), and
$\Pes \lesssim \Peb$ additionally, that is, $\nu/\delta \lesssim 1$
according to \eq{eq:ratioPe}. But then, the most interesting finding
of our analysis is the existence of a regime of renormalized Fick's
law that would eventually be observed
as the probed length scale is increased, see Figs.~\ref{fig:Deff}
and~\ref{fig:Peplane}:
the big particle dynamics would appear to follow pure Fick's law on
small scales ($k^{-1}<\Lb$), anomalous collective behavior on
intermediate scales ($\Lb < k^{-1} < \Ls$), and Fick's law again, but
renormalized, on large scales ($\Ls < k^{-1}$).

To get a sense of the relevant scales, we apply the Stokes--Einstein
relation for the single particle diffusivity,
$\Dvar{(b,s)} = k_B T/6\pi\eta R_\mathrm{(b,s)}$, and introduce the
packing fractions,
$\phi_\mathrm{(b,s)} := \pi R_\mathrm{(b,s)}^2 \dens_\mathrm{(b,s)}$,
of big and small spherical particles of radii $\rb$ and $\rs$,
respectively --- this latter magnitude is more handy at quantifying
how well a collection of particles can be approximated as an ideal
gas. Correspondingly, one gets the dimensionless numbers
\begin{equation}
  \label{eq:peclet2}
  \Peb \approx \frac{\phib}{k \rb},
  \qquad
  \frac{\Pes}{\Peb} \approx \frac{\phis}{\phib} \frac{\rb}{\rs} ,
\end{equation}
which notably are independent of the fluid viscosity $\eta$. If one
takes, e.g., $\phib=0.1$ as in the experiment reported in
\rcites{LRW95,LCXZ14}, the constraints on $\Peb$ and $\Pes/\Peb$ for
observing a big particle dynamics modified by HI yield thresholds on
the probed length scale, $k\lesssim (10 \rb)^{-1}$, and the packing
fraction of the small particles, $\phis \lesssim \rs/10\rb$; the
condition of strong dissimilarity, $\delta=\Ds/\Db = \rb/\rs \gg1$,
translates therefore into a very low packing fraction of small
particles.

A typical experimental realization of the monolayer configuration
consists in trapping colloidal particles at a fluid interface. The
strong dissimilarity limit can then be viewed as a model of
interfacial contamination by surfactant molecules
($\rs\sim 10^{-3}\,\mu\mathrm{m}$), which must be insoluble in the
fluid phases so as to obey the conservation law~(\ref{eq:contsmall}).
In this case, $\Pes$ can be read as the surfactant Marangoni number
\cite{MaSq20}, and \fref{fig:Peplane} captures the effect of
contamination (``small particles'') on the observed evolution of the
colloid distribution (``big particles''). In this context, the
anomalous collective diffusion for submicron--sized particles reported
in \rcites{LRW95,LCXZ14} was probed at $\Peb\sim 1$ (see
\rcite{BDO15}) and \fref{fig:Peplane} allows one to infer $\Pes < 1$
for that experiment. In the opposite limit, the experiment reported in
\rcite{MCDC21} addresses an extremely dilute monolayer
($\phib\sim 10^{-3}$) of micron--sized particles: this is a
configuration with $\Peb \ll 1$ that would already preclude, even in
the unmixed case (no contamination at all), the observation of any
deviation from the pure Fick's law for the colloid. The focus instead
was the measurement of the in-plane flow field $\bu(\br)$ and, because
$\Pes\gg\Peb$ at the tiniest contamination levels reported
($\phis\gtrsim 10^{-3}$ \cite{MCDC21}), the measurements show an
effectively 2D incompressible flow, which is associated to subleading
corrections to the Oseen flow~(\ref{eq:flow2bin}) that we have used in
the model. In effect, this is an indirect evidence of the adiabatic
solution~(\ref{eq:nsSurfAdia}), which must then be viewed as the
manifestation of the long--range ``compressible'' HI between small and
big particles in this dynamic regime. Nevertheless, the model predicts
(see Figs.~\ref{fig:Deff} and \ref{fig:Peplane}) that, upon reducing
the contaminant concentration $\nu$ gradually while keeping fixed
$\delta\gg 1$ and $\Peb>1$ (and thus also the probing scale $k^{-1}$),
the effective dynamics should be observed to crossover from the pure
Fick's law to a renormalized (enhanced) Fick's law first, and to its
breakdown later.

The regime of enhanced Fick's law identified in this model bears
resemblance with the recent report \cite{FCDD26} of a regularization
of the anomalous collective diffusion observed in numerical
simulations and caused by the no-slip constraint enforced by a wall
near the monolayer. There is indeed a close mathematical analogy
between these two scenarios: concerning the HI, the wall is equivalent
to an image monolayer, which suppresses the longest ranged component
of the flow sourced by the monolayer on length scales larger than the
monolayer height $h$ above the wall. Effectively, the image monolayer
would play a role akin to the adiabatically enslaved field
$\nhs^\mathrm{(adia)}$ in \eq{eq:flow2bin}, and the quotient $h/\Lb$
would be equivalent to the ratio $\Ls/\Lb = \delta/\nu$ in
\Fref{fig:Deff} (compare with Fig.~3c in \rcite{FCDD26}).

In summary, we have studied the dynamics of a mixture of colloidal
particles when dominated by the ``compressible'' HI in the monolayer
configuration, extending the results of the single--particle
colloid and providing useful insights. In the limit of identical
particles, the result confirms a previous report \cite{PBPX18} that,
in spite of the anomaly in collective diffusion, the self-diffusion
remains normal. In the opposite limit of very dissimilar particles, we
have focused on the effective dynamics of the big particles and
identified a transitional regime (green region in \fref{fig:Peplane})
between anomalous collective diffusion (blue area) and the absence of
any effect at all by the long--range ``compressible'' hydrodynamic
interaction (red area) --- this absence is only \emph{apparent}
because the effective collective diffusivity of the big particles is
actually unable to discriminate the role played by HI for the small
particles in the red area, compare Table~\ref{tab:regimes}.

The results also suggest future lines of research. When the role of HI
cannot be overtly detected in the big particles dynamics, the focus
could be shifted to other observables like, e.g., the correlations
between small and big particles, as done indirectly in
\rcite{MCDC21}. Alternatively, one could consider changing the sources
of the particle fluxes as another way to alter the effective value of
the ratio $\Pes/\Peb$.  Thus, one can think of extending the model
beyond the ideal--gas approximation: for instance, in the scenario of a
monolayer formed at a fluid interface, the capillary interaction due
to the interfacial deformation appears naturally, and these forces can
be experimentally controlled through different properties of the
particles (weight, geometrical shape, or wetting behavior). Or else,
one can introduce a differential external forcing that drives the
system to a stationary but nonequilibrium state, e.g., a shear flow,
in order to probe rheological properties of the monolayer mixture.

\begin{acknowledgments}
  \label{Acknowledgments}
  The authors acknowledge financial support through Grant
  No.~ProyExcel\_00505 funded by Junta de Andaluc{\'i}a, and Grants
  No.~PID2021-126348NB-I00 and No.~PID2024-156257NB-C21 funded by
  MICIU/AEI/10.13039/501100011033 and by “ERDF/EU”.
\end{acknowledgments}

\appendix

\section{Lack of variational formulation for a nonideal system}
\label{app:noF}

The existence of a variational formulation like expressed by
\eqs{eq:functionalFormulation} is not possible as soon as direct
interactions between the particles are considered. In the generic
case, the induced flow~(\ref{eq:flow}) can be written as
\begin{equation}
  \label{eq:genericflow}
  \bu(\br) = \int d^2\br'\; \mathbf{f}(\br') \cdot \mathsf{O}(\br-\br') ,
\end{equation}
where the thermodynamic force density $\mathbf{f}(\br)$ acting on the
particles can be expressed in the local equilibrium approximation as
\begin{equation}
  \label{eq:f}
  \mathbf{f} = - \dens \nabla \mu = - \nabla p ,
\end{equation}
in terms of the chemical potential $\mu(\dens)$ and the associated
equation of state $p(\dens)$ for the monolayer pressure (we have
applied the thermodynamic relationship $\dens \, d\mu=dp$ valid under
the assumed isothermal conditions).

By introducing a (possibly density--dependent) particle mobility
$\Gamma(\dens)$, the current density in \eq{eq:cont} would be written
down as
\begin{equation}
  \label{eq:j2}
  \bj = \Gamma(\dens) \mathbf{f}(\dens) + \dens \bu
  = - \Gamma(\dens) \nabla p(\dens) + \dens \bu
\end{equation}
Assume the existence of a functional
\begin{equation}
  \label{eq:freeF2}
  \mathcal{F}[n] = \mathcal{F}_\mathrm{local}[n] +
  \mathcal{F}_\mathrm{hydro}[n] 
\end{equation}
that provides the current density~(\ref{eq:j2}) according to
\eq{eq:newj}. The local contribution, which accounts for the first
summand in \eq{eq:j2}, can be easily found due to the local dependence
on $\dens$:
\begin{equation}
  \label{eq:Flocal}
  \mathcal{F}_\mathrm{local}[n] := \int d^2\br\;
  f_\mathrm{local}(\dens(\br)) ,
\end{equation}
with a ``free energy'' density $f_\mathrm{local}(\dens)$ that obeys
the relation
\begin{equation}
  \label{eq:densflocal}
  \frac{d^2 f_\mathrm{local}}{dn^2}
  = \frac{\Gamma(n)}{n} \frac{d p}{d n} .
\end{equation}
The hydrodynamic contribution, on the other hand, must satisfy (see
Eqs.~(\ref{eq:genericflow}, \ref{eq:f}))
\begin{eqnarray}
  \nabla \frac{\delta \mathcal{F}_\mathrm{hydro}}{\delta\dens(\br)}
  & =
  & \mbox{} - \bu(\br) = \int d^2\br'\; \left[\nabla' p(\dens(\br'))\right]\cdot
    \mathsf{O}(\br-\br')
    \nonumber
  \\
  \label{eq:hydrou}
  & =
  & \nabla \int d^2\br'\; \frac{p(\dens(\br'))}{8\pi\eta |\br-\br'|} ,
\end{eqnarray}
where the second line follows from integrating by parts and using
\eq{eq:divOseen}. This result shows that the in-plane compressible
flow is indeed a 2D conservative field, that can be derived from a
potential of the form
\begin{equation}
  \frac{\delta \mathcal{F}_\mathrm{hydro}}{\delta\dens(\br)}
  = \int d^2\br'\; \frac{p(\dens(\br'))}{8\pi\eta |\br-\br'|} .
\end{equation}
Upon taking a second functional derivative, one gets
\begin{equation}
  \frac{\delta^2 \mathcal{F}_\mathrm{hydro}}{\delta\dens(\br)\, \delta\dens(\br')}
  = \frac{1}{8\pi\eta |\br-\br'|} \, \left.\frac{d p}{d \dens} \right|_{\dens=\dens(\br')}.
\end{equation}
The left--hand side of this equation is symmetrical under the exchange
of $\br$ and $\br'$. The right--hand side satisfies this condition
only if $dp/d\dens$ is independent of $\dens$, i.e., if the equation
of state is $p(\dens)\propto\dens$: but this corresponds to an ideal
gas, in which case $dp/d\dens = k_B T$, and
$\mathcal{F}_\mathrm{hydro}[n]$ leads to the field $\Psi$ defined by
\eq{eq:Psi}. This result is a consequence of the nonreciprocal nature
of the HI, a feature reflected, at this level of modelling, in
that the HI are coupled to the density field $\dens(\br)$, see
\eq{eq:cont}, but their source is the pressure field $p(n(\br))$, see
\eq{eq:hydrou}.

%

\end{document}